\documentclass[10pt,conference]{IEEEtran}
\IEEEoverridecommandlockouts
\usepackage{cite}
\usepackage{amsmath,amssymb,amsfonts}
\usepackage{algorithmic}
\usepackage{braket}
\usepackage{fancyhdr}
\usepackage{graphicx}
\usepackage{hyperref}
\usepackage{textcomp}
\usepackage{xcolor}
\def\BibTeX{{\rm B\kern-.05em{\sc i\kern-.025em b}\kern-.08em
    T\kern-.1667em\lower.7ex\hbox{E}\kern-.125emX}}

\fancypagestyle{specialfooter}{%
\fancyhf{}

\fancyfoot[R]{ \noindent\fbox{%
\parbox{\textwidth}{%
{\footnotesize © 2024 IEEE.  Personal use of this material is permitted.  Permission from IEEE must be obtained for all other uses, in any current or future media, including reprinting/republishing this material for advertising or promotional purposes, creating new collective works, for resale or redistribution to servers or lists, or reuse of any copyrighted component of this work in other works.}
}
}}
}

\begin{document}

\title{Quantum Multi-Agent Reinforcement Learning for Aerial Ad-hoc Networks
\thanks{The research is part of the Munich Quantum Valley,
which is supported by the Bavarian state government
with funds from the Hightech Agenda Bayern Plus.} 
}

\makeatletter
\newcommand{\newlineauthors}{%
  \end{@IEEEauthorhalign}\hfill\mbox{}\par
  \mbox{}\hfill\begin{@IEEEauthorhalign}
}
\makeatother

\author{
\IEEEauthorblockN{Theodora-Augustina Dr\u{a}gan}
\IEEEauthorblockA{
\textit{Fraunhofer Institute for Cognitive Systems IKS}\\
Munich, Germany \\
theodora-augustina.dragan@iks.fraunhofer.de}
\and
\IEEEauthorblockN{Akshat Tandon}
\IEEEauthorblockA{
\textit{Airbus Central Research \& Technology}\\
Ottobrunn, Germany \\
akshat.tandon@airbus.com}
\newlineauthors
\IEEEauthorblockN{Carsten Strobel}
\IEEEauthorblockA{
\textit{Airbus Central Research \& Technology}\\
Ottobrunn, Germany \\
carsten.strobel@airbus.com}
\and
\IEEEauthorblockN{Jasper Simon Krauser}
\IEEEauthorblockA{
\textit{Airbus Central Research \& Technology}\\
Ottobrunn, Germany \\
jasper.krauser@airbus.com} \\ 
\newlineauthors
\IEEEauthorblockN{Jeanette Miriam Lorenz}
\IEEEauthorblockA{
\textit{Fraunhofer Institute for Cognitive Systems IKS}\\
Munich, Germany \\
jeanette.miriam.lorenz@iks.fraunhofer.de}
}

\maketitle\thispagestyle{specialfooter}

\begin{abstract}
Quantum machine learning (QML) as combination of quantum computing with machine learning (ML) is a promising direction to explore, in particular due to the advances in realizing quantum computers and the hoped-for quantum advantage. A field within QML that is only little approached is quantum multi-agent reinforcement learning (QMARL), despite having shown to be potentially attractive for addressing industrial applications such as factory management, cellular access and mobility cooperation. This paper presents an aerial communication use case and introduces a hybrid quantum-classical (HQC) ML algorithm to solve it. This use case intends to increase the connectivity of flying ad-hoc networks and is solved by an HQC multi-agent proximal policy optimization algorithm in which the core of the centralized critic is replaced with a data reuploading variational quantum circuit. Results show a slight increase in performance for the quantum-enhanced solution with respect to a comparable classical algorithm, earlier reaching convergence, as well as the scalability of such a solution: an increase in the size of the ansatz, and thus also in the number of trainable parameters, leading to better outcomes. These promising results show the potential of QMARL to industrially-relevant complex use cases. 
\end{abstract}


\section{Introduction}
\label{sec:introduction}
In the field of aerospace communication, technology has already enabled wireless mobile nodes to connect to each other and to act as both relay points and access points. This allows the creation of flying ad-hoc networks (FANET). Architectural advancements have recently been made in this field, such as free-space optical communication (FSO) hardware, as well as the corresponding communication management software~\cite{helle2022agentbased, helle2022decentralized}. This means that the FANETs, which were usually made up of unmanned aerial vehicles (UAV), can now be formed by commercial aircrafts, satellites, as well as by other platforms, enabling them to exchange information. The main challenges of FANETs, when compared to other types of ad-hoc networks, are the high mobility degree and the low node density, which renders link disconnections and network partitions more likely~\cite{muhammad2020fanets}.

The FANET nodes can therefore collaborate to overcome the connectivity challenge by addressing it as a common goal. Each node can choose which other nodes to open a communication channel with, such that as many nodes as possible are directly or indirectly reachable by the rest of the network. There are several benefits for aircrafts to create ad-hoc networks that motivate this work, such as for passenger and aircraft connectivity, as well as for acting as a backbone for internet service providers. For this purpose, a centralized decision-making process would be able to apply fully-informed routing protocols and dynamically adjust connections as topology changes. While such strategies perform better than a collection of random agents, they are impractical in FANETs: they does not scale well with a large number of network nodes and become impractical, and thus decentralized solutions are preferable \cite{muhammad2020fanets, helle2022decentralized, kim2023drones}. 

Multi-agent reinforcement learning (MARL) is a collection of methods designed for multi-agent systems (MAS). They assume that each agent is a different entity which can learn how to behave in an environment by interacting with it. It usually entails two processes: training, when the agents update their internal rules depending on the feedback caused by their actions, and execution, when they act according to those rules. MARL could provide here a solution, as it contains algorithms where the agents could use global information during training, and only local information during execution. The advantage of these methods is the reduction in inter-agent communication overhead. However, this paradigm comes with certain drawbacks, such as the poor scalability, a high demand of computational resources, as well as only having partial access to environmental information. Therefore, we explore if a quantum-enhanced MARL (QMARL) could help to tackle some of these issues and could lead to a better performance of the agents.

The contributions detailed in this work are:
\begin{itemize}
    \item We present an HQC multi-agent proximal policy optimization algorithm, where the core of the centralized critic is a data reuploading variational quantum circuit (VQC). The VQC is designed so that it is compatible with the quantum technology currently available.
    \item We model an aerial communication use case against which both the aforementioned HQC MARL algorithm and its classical counterpart are benchmarked. 
    \item We scale up the size of the VQC with respect to the number of layers and, respectively, the complexity of the use case, and assessed the scalability of our solution. We also characterize the VQC using two quantum metrics that are well-motivated by literature, namely expressibility and entanglement capability. The purpose is to observe whether any correlation could be drawn between the performance of the HQC solution and the embedded quantum module.
\end{itemize}

This paper is structured as follows: the next section is a dive into the theoretical basis notions of MARL, followed by a presentation of the current state of the art in QMARL. The fourth section presents the MARL environment, therefore the task at hand, while section~\ref{sec:algorithm} details the  classical MARL algorithm the solution is built on and the process of embedding a quantum kernel into the training process. In section~\ref{sec:evaluation} we introduce the methods for evaluating the classical and quantum solutions with respect to their performance, as well as to their architectural properties. In section~\ref{sec:results} we present the results of the QMARL solution and then draw the conclusions in the final chapter. 

\section{Background}
\label{sec:background}
In this section, we will introduce the (MA)RL paradigm and its applications, as well as the main challenges encountered in the development of such algorithms and the main categories in which they are divided. Finally, we present the method we chose to build our QMARL algorithm on.

MARL is a collection of methods which make use of the reinforcement learning (RL) paradigm in order to enable agents to successfully behave in MASs. While supervised and unsupervised ML propose training a model on input data in order to perform a task, RL agents interact with their environment and observe the feedback they get as reward in order to improve their behaviour in the environment and obtain better rewards. These methods applied to MAS contexts can achieve results comparable to professional human players in video games~\cite{ellis2023smacv2}, as well as perform well on industrially-relevant use cases such as smart manufacturing~\cite{bahrpeyma2022review}, UAV cooperation for network connectivity and path planning~\cite{qie2019joint}, and energy scheduling of residential microgrid~\cite{fang2019multi}.

The ubiquity of MAS and extensive research of RL methods motivated the development of existing single-agent RL algorithms into MARL solutions. However, this yielded new challenges: since the state of an environment does not depend on the actions of a single agent, the environment is thus non-stationary with respect to that agent. Scalability and the curse of dimensionality are also characteristics of MARL, since the dimensions of the joint state and action spaces can steeply increase and thus make solutions demand more computational resources. Finally, most environments are only partially observable for each agent, while RL algorithms assume the agent has full knowledge of the environment.

In a MARL solution there are two stages, training, when the model of the behaviour of each agent is updated through interactions with the environment, and execution, when a trained model starts performing its assigned task in the environment. Depending on whether information is shared between the agents during each of these two stages, three approaches can be distinguished:
\begin{itemize}
    \item Centralized training, centralized execution (CTCE): agents are always able to communicate and can be viewed as one single agent. The drawback of this approach is that agents are expected to exchange information during execution, which decreases scalability and increases overhead.
    \item Decentralized training, decentralized execution (DTDE): agents never communicate and act as independent RL single agents. While this option has little overhead in both the development and the testing of the solutions, it also underperforms when compared to other approaches.
    \item Centralized training, decentralized execution (CTDE):  agents are able to communicate during the training process, for example by having access to simulator information or by communicating through a network. During execution, information is not shared anymore.
\end{itemize}
We chose to implement an algorithm of the CTDE approach, since this paradigm is able to help mitigate the scalability and the partial observability issues. Since knowledge sharing only happens during training, agents may learn better than by only having local information, but they also avoid the informational exchange overhead during execution, where they act as single agents.

\section{Related Works}
\label{sec:related}
This section provides an introduction into the present advancements in the field of QMARL. We start with a general presentation of quantum methods in ML and in RL, and then present the possible paths of development of quantum-enhanced solutions. We then conclude with a presentation of the current status of QMARL approaches through selected works.

Quantum machine learning is a collection of methods that can be found at the intersection between quantum computing and machine learning. In this work, we understand it as using quantum phenomena such as superposition, entanglement, and inference in order to gain a computational advantage or a better performance on applications where input data is classical. The motivation behind this field is the fact that methods with quantum modules were shown to have lower time complexities~\cite{wiebe2014quantum, li2022quantum, lloyd2013quantum}, better performances with respect to the application-specific metrics~\cite{ullah2022cardiopathy, abbas2021power}, as well as theoretical advantages, such as a better generalization in cases where data samples are limited~\cite{caro2022generalization}.

These aspects also apply to quantum reinforcement learning (QRL), where several works already proposed multiple directions~\cite{meyer2024survey}. These can be divided into four main pillars: quantum-inspired methods(classical algorithms that mimic quantum principles), VQC-based function approximators, RL algorithms with quantum methods, and fully-quantum RL. The second category comprises the only algorithms with quantum modules that are suitable for the currently available quantum hardware, also known as noisy intermediate-scale quantum (NISQ) devices~\cite{Preskill_2018}. The VQC-based subdomain contains classical RL algorithms that originally use neural networks (NN) as function approximators and now replaced them with VQCs. Such solutions were already proposed for use cases such as robotics~\cite{heimann2022quantum}, wireless communication~\cite{chen2020vqcdrl}, optimization~\cite{skolik2023robustness}, and logistics~\cite{correll2023quantum}. In such works, VQCs can be employed in order to compute the suitability of an environmental state, the probabilities of an action to be taken in a given state, or other intermediary computations that help the agent to successfully navigate the environment.

Most of the QMARL literature also focuses on these VQC-based NISQ-friendly algorithms. For example, an actor-critic QMARL algorithm was applied on two cooperative tasks: smart factory management and mobile access generated by UAVs~\cite{park2023quantum, yun2023quantum}. Three types of solutions were proposed, depending on the implementation of the actor and, respectively, of the centralized critic: entirely quantum (QQ), a quantum-centralized critic and classical actors (QC), and entirely classical (CC). The VQCs of the QQ and QC solutions consisted of an angle data encoding and a trainable layer of rotational gates and CNOT entanglement gates. Results show that the architecture of quantum actors and a quantum critic learnt more efficiently than other approaches~\cite{park2023quantum, yun2023quantum}. For comparable rewards to be achieved during training, the classical approach would require two orders of magnitude more trainable parameters. Moreover, if projection value measure is used for dimensionality reduction on the action space of the quantum solution, it scales better than other classical algorithms once the action space reaches the order of $2^{16}$. This hints towards a better suitability of QMARL solutions for industrially-relevant MAS use cases, when compared to classical MARL. 

A similar work makes use of quantum actors and a quantum centralized critic in a realistic decentralized environment of multi-UAV cooperation in the presence of noise~\cite{park2023quantummulti}. The actions of the UAVs in that use case are their movements, which should conduct to a better-performing UAV network as observed by the end users on the ground. The simulation environment is challenged through noise: generalised Cauchy state value noise and Weibull distribution-like noise on the action values, which render the simulation environment closer to a real use case. The presence of environmental and action noise is actually favorable for the QMARL solutions, which then converge faster and to higher rewards than their noiseless or classical counterparts.

Another hybridised paradigm present in literature is evolutionary optimization, in which the optimization of the parameter set of a model is done analogously to natural selection. Several initial sets of potential parameters are generated and then, in an iterative process, the best candidates are selected based on a fitness function. New candidate parameters are generated, until a satisfactory set of parameters is achieved. Such an optimization process can be employed to train the embedded VQC in a QMARL model to solve a coin game in which both the state space and the actions taken are discrete~\cite{kolle2024multiagent}: in a grid-like environment two agents compete against each other in order to maximize the number of coins collected. Multiple evolution strategies were applied to the QMARL algorithm and were benchmarked against similar solutions which employ instead NNs. Results show that quantum-inspired methods are able to reach comparable results to classical ones, while reducing the parameter count to half.

\section{Environment}
\label{sec:environment}
\begin{figure}[htbp]
\centerline{\includegraphics[width=0.45\textwidth]{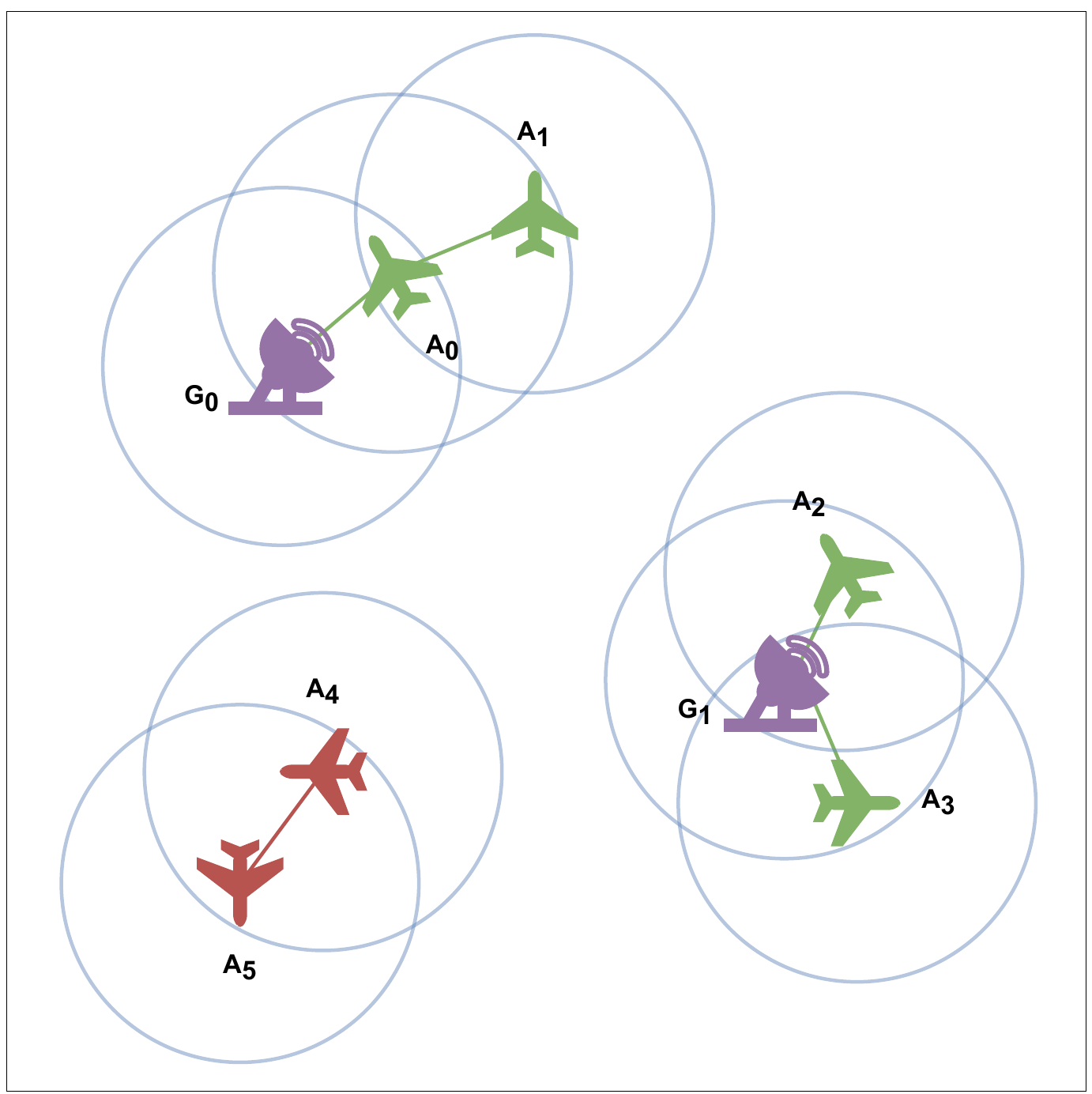}}
\caption{An environment of $N = 8$ entities: $N_A = 6$ aircrafts and $N_G = 2$ ground stations.}
\label{fig:env_example}
\end{figure}

To address inter-plane communication via both MARL and QMARL algorithms, an environment to simulate the aircrafts and ground stations needs to be defined. This section introduces such an environment from two points of view: the physical simulation of the environment, as well as its mathematical formalisation as a partially-observable Markov decision process.

The environment is a simulated MAS of several entities, where an entity is either an aircraft or a ground station. For each entity, its initial positions and constant velocities on the $x$ and $y$ axes are randomly and uniformly generated, with the velocities of the ground stations being $0$. Time is discretized into time steps and at each time step the agents move according to their velocities. Afterwards, they decide who to connect to, as each of them is able to connect to maximally $2$ entities. If both agents decided to connect to each other, the connection is established, else not. The goal of the agents is to take good connection decisions and create local ad-hoc networks such that a maximally achievable number of aircrafts is connected to the ground.

There are in total $N = N_A + N_G$ entities, where $N_A$ is the number of aircrafts and $N_G$ is the number of ground stations. An aircraft is connected to the ground as long as it has an uninterrupted (multi-hop) link to a ground station. For example, in the environmental state shown in Fig.~\ref{fig:env_example}, aircrafts $A_0$, $A_2$ and $A_3$ are connected directly to the ground stations $G_0$ and $G_1$, whereas $A_1$ is connected indirectly through $A_0$. Aircrafts $A_4$ and $A_5$ are connected to each other, but as no other aircrafts or ground stations are in range, they have no access to communication (where ranges are represented through blue circles). A simulation is run for $T = 50$ time steps, and the goal of each aircraft is to properly choose to which other aircrafts to connect in order to maximize the total number of aircrafts connected to the ground.

The environment can be modelled as a decentralized  partially observed Markov decision process (Dec-POMDP)~\cite{oliehoek2016concise} denoted as $\mathcal{M} = (\mathcal{D}, \mathcal{S}, \mathcal{A}, \mathcal{O}, R, T)$. In this notation, $\mathcal{D} = \{1,2,\dots,N_A\}$ is the set of agents, $\mathcal{S}$ is the set of states, $\mathcal{A}$ is the set of joint actions, $\mathcal{O}$ is the set of observations, $R$ is the immediate reward function and $T$ is the problem horizon.

In the following notations, all values correspond to the properties of the environment at time step $t$, but the index $t$ is omitted for clarity purposes. The state of the environment $\mathcal{S} = {x_{e_i},y_{e_i},v_{x_{e_i}},v_{y_{e_i}}}_{1 \leq i \leq N}$ contains the $x$ and $y$ axis positions and the velocities of all entities $\{e_i\}_{1 \leq i \leq N}$. The environment state $\mathcal{S}$ is not visible to any of the entities, to reflect the real-world application of such an environment.

The joint action set is $\mathcal{A} = \{ a_{a_i} \}_{1 \leq i \leq N_A}$, where the action $a_{a_i}$ of each aircraft $a_i$ is defined as:
\begin{equation}
a_{a_i} = \{ c_{e_0}, c_{e_1},\dots,c_{e_N}\},
\label{eq:env_act}
\end{equation}
where $c_k \in (0, 1)$ is a value directly proportionate to how desirable  the connectivity with entity $e_k \neq a_i$ is to the aircraft $a_i$ and the connectivity choice corresponds to the highest $2$ values.

The joint observation set is $\mathcal{O} = \{ o_{a_i} \}_{1 \leq i \leq N_A}$, where the observation $o_{a_i}$ of each aircraft $a_i$ is defined as:
\begin{equation}
o_{a_i} = \{\text{ptg}_{a_i}, \text{ptg}_{e_1}, \text{lk}_{e_1}, \text{oc}_{e_1}, \dots, \text{ptg}_{N-1}, \text{lk}_{N-1}, \text{oc}_{N-1}\},
\label{eq:env_state}
\end{equation}
where $\text{ptg}_{e_k} = 1$ if the entity $e_k \neq a_i$ has a path to the ground and $\text{ptg}_{e_k} = 0$ otherwise. The normalized link range $\text{lk}_{e_k} \in [0, 1]$ shows for how many steps, out of the total number of simulated environmental time steps, aircraft $a_i$ and entity $e_k$ will be in reach of each other. If they are currently not in range, $\text{lk}_{e_k} = -1 $. Finally, the normalized occupied connections variable $\text{oc}_{e_k} \in [-1, 1]$ indicates how many of the maximally available connections are occupied. If $\text{oc}_{e_k} = -1$, entity $e_k$ has no active connections, and if $\text{oc}_{e_k} = 1$ , it reached the maximal number of simultaneous connections, which is set at two for the use case scenarios tackled in this work.

The reward for each agent is chosen as a global reward $R$:
\begin{equation}
R = \frac{1}{N_A} \sum_{i=1}^{N_A} \text{ptg}_i,
\label{eq:env_rew}
\end{equation}
which is the averaged path to ground of all aircrafts at a given time step $t$.

\section{Algorithm}
\label{sec:algorithm}
This chapter details the QMARL algorithm that solves the environment defined in the previous section. It is based on the multi-agent proximal policy optimization (MAPPO) algorithm. The implementation was adapted from the MARLLib library~\cite{hu2023marllib} and benchmarked against its classical counterpart, both following the original MAPPO algorithm~\cite{yu2022surprising}. 

The MAPPO algorithm is the multi-agent version of the  proximal policy optimization (PPO) RL algorithm~\cite{schulman2017proximal}, which is widely used in literature due to its performance on complex use cases, such as robotics~\cite{moon2022path} and video games~\cite{openai2019dota}. Like other actor-critic RL algorithms, it uses two function approximators in order to compute the next best action to be taken by the agent. The actor, also known as the policy function, outputs the probabilities of each action to be taken in a state. The critic, also known as the value function, estimates the value of a given state of the environment, directly proportional to the expected reward to be obtained during the episode from that state onwards. These two function approximators are usually implemented as NNs, in order to accommodate for state and action spaces of high dimensions. The main improvement brought by PPO in the actor-critic family is using trust region policy updates with first-order methods, as well as clipping the objective function. This enables the method to be more general than other trust region policy methods and have a lower sample complexity~\cite{schulman2017proximal}.

The MAPPO maintains the same architecture of the PPO, with two types of NNs: the individual policy $\pi_\theta$ (actor) of each agent and the collective value function $V_\phi(O)$ (critic), where $O$ is the global environmental observation of the Dec-POMDP. The final goal of our solution is to maximize the mean path to ground at each time step, reflected by minimizing the cumulative reward (CR) of all agents during an episode:
\begin{equation}
    \text{CR} = T*N_A*R.
\label{eq:cr}
\end{equation}
In order to achieve this, the MAPPO algorithm minimizes two losses through two Adam optimizers~\cite{kingma2017adam}, during the same training process~\cite{yu2022surprising}. The loss that the actor network will minimize during training is:
\begin{equation}
L(\theta) = \frac{1}{Bn} \sum_{i=1}^B \sum_{k=1}^n \left( a_{\theta,i}^{(k)} +  \sigma S[\pi_\theta(o_i^{(k)})]\right),
\end{equation}
\label{eq:actor_loss}
where $a_{\theta,i}^{(k)} = \text{min}(r_{\theta,i}^{(k)} A_{i}^{(k)},\text{clip}(r_{\theta,i}^{(k)}, 1-\epsilon, 1+\epsilon) A_{i}^{(k)})$ is the PPO-specific clipped advantage function $A$, which can be understood as an estimated relative value function. Furthermore, $\theta$ is the parameter set of the actor network, $B$ is the batch size, $n$ is the number of agents, S is the policy entropy, $\sigma$ is the entropy coefficient hyperparameter, and $A_i^{(k)}$ is the advantage function.

The loss of the centralized critic is:
\begin{equation}
L(\phi) = \frac{1}{Bn} \sum_{i=1}^B \sum_{k=1}^n \text{max}((V_\phi(o_i^{(k)}) - \hat{R_i})^2, ( v_{\phi,i}^{(k)} - \hat{R_i})^2).
\end{equation}
\label{eq:critic_loss}
In this case the clipped objective is the clipped value function $v_{\phi,i}^{(k)} =  \text{clip}(V_\phi(o_i^{(k)}), V_{\phi_{old}}(o_i^{(k)}) - \epsilon, V_{\phi_{old}}(o_i^{(k)}) + \epsilon)$, $\phi$ is the parameter set of the critic network and $\hat{R_i} = \gamma \cdot \text{CR} $ is the discounted cumulative reward. The values chosen for the MAPPO hyperparameters in our implementation are found in Table~\ref{tab:hyperparams}.
\begin{table}[htbp]
\caption{Hyperparameter values}
\begin{center}
\begin{tabular}{|c|c|}
\hline
\textbf{Hyperparameter}&\textbf{Value} \\
\hline
GAE discount factor $(\lambda_\text{GAE})$ & 0.99  \\
entropy factor ($\epsilon$) & 0.2  \\
clipping factor ($\sigma$) & 0.01  \\
KL penalty & 0.2  \\
learning rate & 0.0001  \\
reward discount factor ($\gamma$) & 0.99  \\
\hline
\end{tabular}
\label{tab:hyperparams}
\end{center}
\end{table}

\subsection{Quantum Module}

\begin{figure*}[htbp]
\centerline{\includegraphics[width=\textwidth]{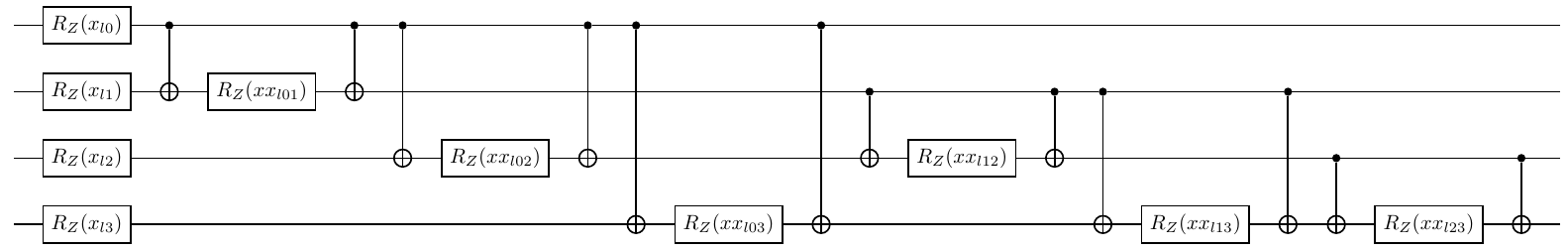}}
\caption{Feature map (FM) of the variational quantum circuit.}
\label{fig:feature_map}
\end{figure*}

\begin{figure}[htbp]
\centerline{\includegraphics[width=0.48\textwidth]{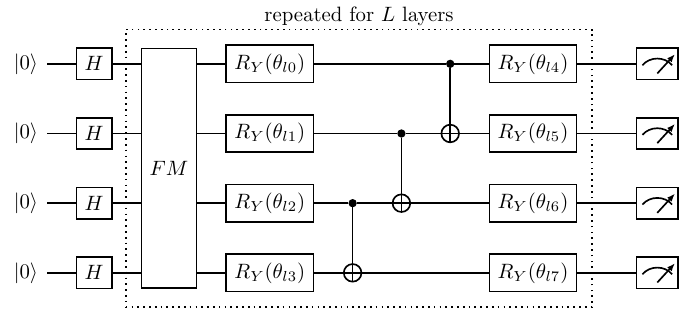}}
\caption{Structure of the VQC core of the centralized critic, where FM is the feature map presented in Fig.~\ref{fig:feature_map}.}
\label{fig:vqc_ansatz}
\end{figure}

The hybrid quantum-classical variant of the MAPPO (QMAPPO) algorithm we employ is obtained by replacing a part of the centralized critic NN with a VQC, leaving the rest of the modules and the training policy intact. The critic NN has three parts: the pre-processing block, the core block, and the post-processing block. Each block is formed of fully-connected linear layers followed by the hyperbolic tangent activation function. 

In the case of the QMAPPO solution, the core NN block is replaced by a VQC, whose structure is displayed in Fig.~\ref{fig:vqc_ansatz}. It is a data reuploading quantum circuit of $4$ qubits, which repeats $L$ layers of a feature map (FM) and of a trainable ansatz. The feature map is a second-order Pauli-$Z$ evolution circuit (the $ZZ$ feature map), in which the rotational angles are $x_{lq_i} = f(o_{lq_i} \cdot \xi_{lq_i})$ and $xx_{lq_iq_j} = 2(\pi - x_{lq_i})(\pi - x_{lq_j})$, where $l \in \{0, 1, 2\}$ is the layer index, $q_i, q_j \in \{0, 1, 2, 3\}, q_i < q_j$ are input data indices in a layer ~\ref{fig:feature_map}, $o$ are the pre-processed input features, $\xi$ are trainable input scaling weights, and $f$ is the pre-processing function, which is either the identity or the inverse tangent function.

Depending on whether we repeat the feature map for $L = 1,~2 \text{ or } 3$ layers, we obtain VQC-1, VQC-2 and VQC-3 and embed then $4,~8, \text{ or, respectively, } 12$ features of the pre-processed input and thus the pre-processing linear layer has an output dimension of $4, 8 \text{ or } 12$ as well. When $f$ is the identity function, so no further scaling is applied, the circuits are referred to as VQC-1N, VQC-2N and VQC-3N, and if $f$ is the inverse tangent function, they are referred to as VQC-1A, VQC-2A and VQC-3A. The classical counterpart of each VQC-based solution has a critic core NN block of two hidden layers that have the same number of neurons. For a fair comparison, the number of neurons per layer is chosen such that the total weight count is as similar as possible between the MARL and QMARL solutions, respectively. The classical solutions are denoted as NN-$X$, where $X$ is the number of neurons in a hidden layer.

For the optimization of the actor and critic modules as a whole, the Adam optimizer is used. Nevertheless, the optimization of the weights of the quantum circuit is done classically through the simultaneous perturbation stochastic approximation (SPSA) optimizer~\cite{spall1998overview}, chosen for its efficiency: it needs only three circuit executions, whereas optimizers which use the parameter-shift-rule need $O(2n)$ circuit executions.

\section{Evaluation}
\label{sec:evaluation}

In this section we present the two types of metrics that are used to benchmark all solutions: performance metrics, which indicate how well the agents perform at evaluation during training, as well as architectural metrics which are indicated by literature to give an insight into the learning capability of a quantum-enhanced solution. 
\subsection{Performance Metrics}
In order to evaluate how well each architecture performs, which is how well the agents choose communication links in environments of the same size they were trained on, but of new configurations, we propose the following metrics:
\begin{itemize}
    \item Maximal Cumulative Reward (MCR): the maximal value of the aggregated mean reward during training across all experiments of a given solution, sampled at evaluation;
    \item Converged Cumulative Reward (CCR): the mean value of the aggregated mean reward during training across all experiments of a given solution after $10^6$ time steps of training. This is proposed since after $10^6$ time steps, most solutions have converged to a stable CR, therefore it can be seen as a more robust average of the CR;
    \item Convergence Speed (CS): the number of thousands of time steps it takes for a model to reach an MCR $25\%$ higher than the average CR achieved by random agents (Rand).
\end{itemize}

\subsection{Quantum Metrics}

A significant endeavour in literature is to anticipate the performance of a quantum-enhanced solution and to compare between different solution architectures on the same task~\cite{bowles2024better}. Among these architectural metrics, one may find the trainability~\cite{mcclean2018barren}, the expressibility, the entanglement capability~\cite{sim2019expressibility}, and the normalized effective dimension~\cite{abbas2021power}. Moreover, since most metrics are estimated on sampled sets of the trainable parameters of a VQC and can get computationally demanding, machine learning-based estimating solutions were proposed as well~\cite{aktar2023expressibility}. While clear correlations are still to be found between any proposed metric and the performance of the corresponding VQC-based solutions, two quantum metrics are widely used in literature~\cite{sim2019expressibility} and are presented in the remaining of this chapter: expressibility and entanglement capability.

\subsubsection{Entanglement Capability}

The entanglement capability (Ent) of a VQC is an indicator of how entangled its output states are~\cite{sim2019expressibility}. This metric is based on the Meyer-Wallach (MW) entanglement of a quantum state as follows:
\begin{equation}
    \text{Ent} = \frac{1}{|S|} \sum_{\Theta_i \in S} Q(\psi_i),
    \label{eq:entang}
\end{equation}
where $Q(\psi_i)$ is the MW entanglement applied to the output quantum state $\psi_i$, generated by a sampled vector of parameters $\Theta_i \in S $, where $ S $ is the ensemble of the sampled parameter vectors. The entanglement capability is bounded, $\text{Ent} \in [0, 1]$, and its value is directly proportional to how entangled the output states are. For example e.g., $\text{Ent} = 1$ for a circuit that generates the maximally-entangled Bell states.

\subsubsection{Expressibility}

The expressibility (Expr) of a circuit is a quantum metric that indicates how close the distribution of the output states of that circuit is to the Haar ensemble, an uniform distribution of random states. Therefore, it measures how well a circuit covers the Hilbert space and uses for this purpose the Kullback-Leibler (KL) divergence between the two distributions:
\begin{equation}
    \text{Exp} = \text{D}_{\text{KL}}(P_{\text{VQC}}(F, \Theta)~||~P_{\text{Haar}}(F)),
    \label{eq:express}
\end{equation}
where $P_{\text{PQC}}$ is the estimated probability distribution of the fidelities between pairs of samples of output states of the VQC, $P_{\text{Haar}} = (N - 1)(1 - F)^{N-2}$ is the probability distribution function between states of the Haar ensemble, $N$ is the dimension of the Hilbert space, and $F = |\braket{\psi_\theta | \psi_\phi}|^2$ is the fidelity function between two quantum states $\ket{\psi_\theta}$ and $\ket{\psi_\phi}$.

The quantum metrics of each VQC were computed using the qleet library~\cite{azad2023qleet}, where they are implemented according to the definitions given in this section. In the following section, the results of the classical and QMARL models are introduced and the latter are benchmarked against these two quantum metrics. 

\section{Results}
\label{sec:results}

To assess the scalability of the classical and quantum-enhanced solutions with the complexity of the use case, we benchmark them against two scenarios:
\begin{itemize}
    \item \textbf{4A1S:} A basic scenario of $N = 5$ entities, with $N_A = 4$ aircrafts and $N_G = 1$ ground station. The size of the observation of an agent is $\text{dim}(o) = 13 $ and the action size of an agent is $\text{dim}(a) = 4 $. Therefore, the collective observation space is of size $\text{dim}(O) = 52 $ and the collective action size is $\text{dim}(a) = 16$. The cumulative reward achieved by random agents of uniformly generated actions is $\text{CR}_{\text{Rand}} = 60.20$.
    \item \textbf{5A2S:} A more complex scenario of $N = 7$ entities, with $N_A = 5$ aircrafts and $N_G = 2$ ground stations. The size of the observation of an agent is $\text{dim}(o) = 19 $ and the action size of an agent is $\text{dim}(a) = 6 $. Therefore, the collective observation space is of size $\text{dim}(O) = 95 $ and the collective action size is $\text{dim}(a) = 24$. The cumulative reward achieved by random agents of uniformly generated actions is $\text{CR}_{\text{Rand}} = 84.88$.
\end{itemize}

\begin{figure*}[htbp]
\centerline{\includegraphics[width=0.8\textwidth]{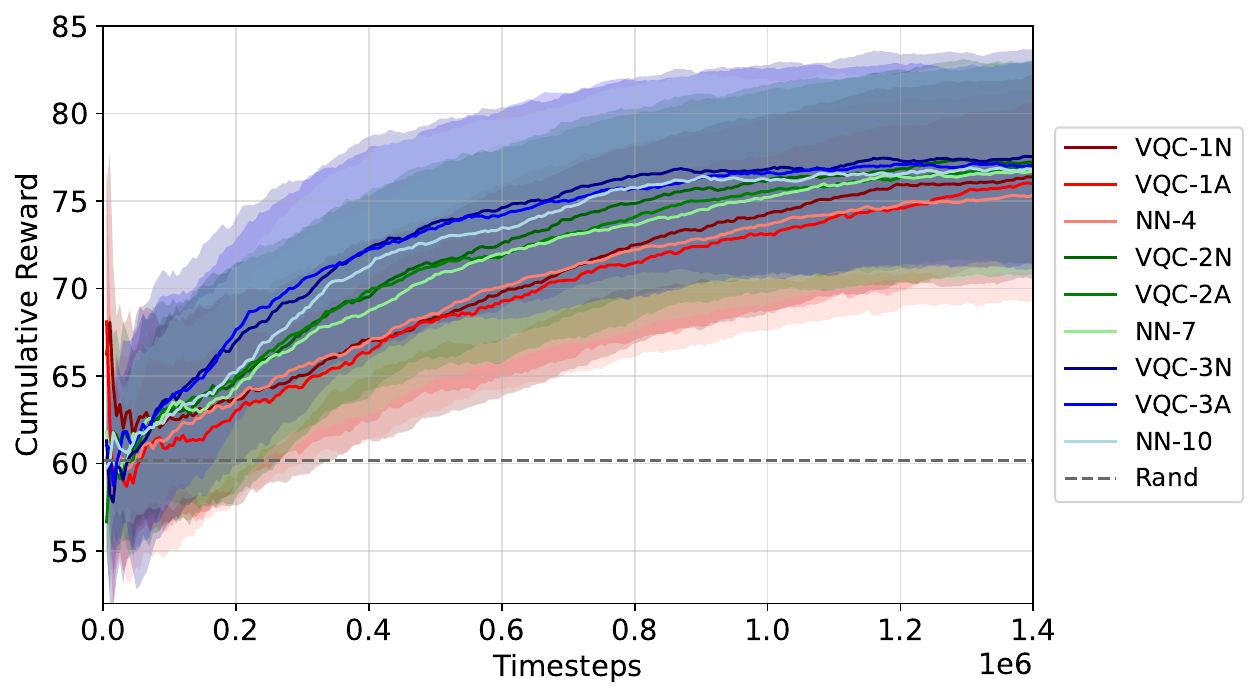}}
\caption{Smoothed aggregated cumulative reward at evaluation of all classical and QMARL solutions in the 4A1S scenario.}
\label{fig:all_results_4a1s}
\end{figure*}

\begin{figure*}[htbp]
\centerline{\includegraphics[width=0.8\textwidth]{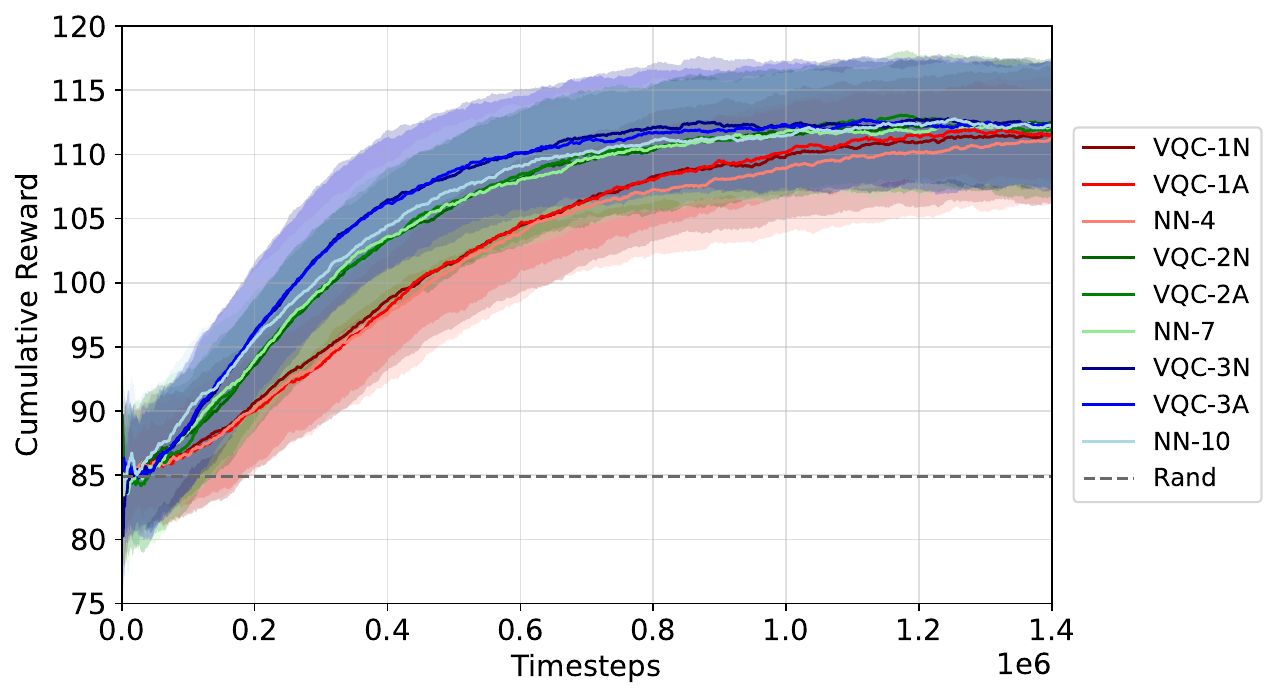}}
\caption{Smoothed aggregated cumulative reward at evaluation of all classical and QMARL solutions in the 5A2S scenario.}
\label{fig:all_results_5a2s}
\end{figure*}

\begin{table*}[htbp]
\caption{The number of classical weights (CW), quantum weights (QW), and total weights (TW) of all solutions in the 4A1S scenario, together with their respective expressibility (Expr) and entanglement capability (Ent), and their performance metrics: maximal cumulative reward (MCR), converged cumulated reward (CCR), and converge speed (CS) in thousands of time steps. }
\begin{center}
\begin{tabular}{|r|c|c|c|c|c|c|c|c|}
\hline
\textbf{Sol} & CW & QW & TW & Expr & Ent & $\text{MCR}$ & $\text{CCR}$ & $\text{CS}$ \\
\hline
NN-4 & 249 & - & 249 & - & - &  $ 84.23 \pm 10.53 $ & $ 76.59 \pm 3.78 $ & 255 \\
VQC-1N & 241 & 12 & 253 & $ 0.0013 \pm 0.0001 $ & $ 0.8476 \pm 0.0084 $ & $ 89.63 \pm \phantom{0}6.26 $ & $ 77.91 \pm 3.90 $ & 335 \\
VQC-1A & 241 & 12 & 253 & $ 0.0030 \pm 0.0004 $ & $ 0.8043 \pm 0.0091 $ & $ 89.93 \pm \phantom{0}0.57 $ & $ 77.16 \pm 5.09 $ & 203  \\
\hline
NN-7 & 447 & - & 447 & - & -& $ 86.56 \pm \phantom{0}1.13 $ & $ 77.90 \pm 3.91 $ & 195 \\
VQC-2N & 453 & 24 & 477 & $ 0.0012 \pm 0.0002 $ & $ 0.8308 \pm 0.0062 $ & $ 90.16 \pm \phantom{0}3.05 $ & $ 78.01 \pm 3.53 $ & 260 \\
VQC-2A & 453 & 24 & 477 & $ 0.0025 \pm 0.0006 $ & $ 0.8128 \pm 0.0091 $ & $ 87.43 \pm \phantom{0}1.58 $ & $ 77.50 \pm 3.91 $ & 141 \\
\hline
 NN-10 & 663 & - & 663 & - & - & $ 87.76 \pm \phantom{0}9.82 $ & $ 77.24 \pm 4.30 $ & 215 \\
VQC-3N & 665 & 36 & 701 & $ 0.0013 \pm 0.0002 $ & $ 0.8278 \pm 0.0072 $ & $ 88.56 \pm \phantom{0}9.15 $ & $ 78.01 \pm 3.95 $ & 180 \\
VQC-3A & 665 & 36 & 701 & $ 0.0025 \pm 0.0005 $ & $ 0.8186 \pm 0.0076 $ & $ 89.76 \pm \phantom{0}6.45 $ & $ 77.73 \pm 4.08 $ & 133 \\
\hline
\end{tabular}
\label{tab:performance_metrics_4a1s}
\end{center}
\end{table*}

\begin{table*}[htbp]
\caption{The number of classical weights (CW), quantum weights (QW), and total weights (TW) of all solutions in the 5A2S scenario, together with their respective expressibility (Expr) and entanglement capability (Ent), and their performance metrics: maximal cumulative reward (MCR), converged cumulated reward (CCR), and converge speed (CS) in thousands of time steps. }
\begin{center}
\begin{tabular}{|r|c|c|c|c|c|c|c|c|}
\hline
\textbf{Sol} & CW & QW & TW & Expr & Ent & $\text{MCR}$ & $\text{CCR}$ & $\text{CS}$ \\
\hline
NN-4 & 433 & - & 433 & - & - & $ 119.93 \pm 1.44 $ & $ 106.56 \pm 5.76 $ & 360 \\
VQC-1N & 425 & 12 & 437 & $ 0.0013 \pm 0.0001 $ & $ 0.8476 \pm 0.0084 $ & $ 119.69 \pm 1.51 $ & $ 107.28 \pm 5.72 $  & 312 \\
VQC-1A & 425 & 12 & 437 & $ 0.0030 \pm 0.0004 $ & $ 0.8043 \pm 0.0091 $ & $ 125.23 \pm 1.78 $ & $ 107.34 \pm 6.55 $  & 246 \\
\hline
NN-8 & 873 & - & 873 & - & - & $ 122.13 \pm 1.60 $ & $ 109.76 \pm 4.49 $ & 210 \\
VQC-2N & 809 & 24 & 833 & $ 0.0012 \pm 0.0002 $ & $ 0.8308 \pm 0.0062 $ & $ 120.76 \pm 5.14 $ & $ 109.81 \pm 4.57 $ & 192 \\
VQC-2A & 809 & 24 & 833 & $ 0.0025 \pm 0.0006 $ & $ 0.8128 \pm 0.0091 $ & $ 121.56 \pm 7.31 $ & $ 109.95 \pm 4.60 $ & 202 \\
\hline
 NN-11 & 1224 & - & 1224 & - & - & $ 121.03 \pm 5.43 $ & $ 110.17 \pm 4.31 $  & 181 \\
VQC-3N & 1193 & 36 & 1229 & $ 0.0013 \pm 0.0002 $ & $ 0.8278 \pm 0.0072 $ & $ 123.29 \pm 7.76 $ & $ 111.02 \pm 4.06 $ &  145 \\
VQC-3A & 1193 & 36 & 1229 & $ 0.0025 \pm 0.0005 $ & $ 0.8186 \pm 0.0076 $ & $ 121.96 \pm 2.30 $ & $ 110.89 \pm 3.93 $  & 186 \\
\hline
\end{tabular}
\label{tab:performance_metrics_5a2s}
\end{center}
\end{table*}

Three experiments are performed for each architecture -- scenario pair. The models are trained for $1\,400\,000$ time steps, where the random seeds of each experiment are $\{0, 1, 2\}$ and the CR is sampled every 1000 time steps. In Fig.~\ref{fig:all_results_4a1s} and in Fig.~\ref{fig:all_results_5a2s} the results are plotted and smoothed using the exponential moving average, with the error bands representing the standard error of the three experiments. Tables~\ref{tab:performance_metrics_4a1s} and~\ref{tab:performance_metrics_5a2s} present the aggregated results for all chosen architectures and, respectively, performance metrics, together with the number of classical, quantum and total trainable weights.

When it comes to the smaller-scale 4A1S scenario, all of the QMAPPO solutions with the inverse tangent input scaling function (VQC-1A, VQC-2A, and VQC-3A) require around half as many iterations to converge to the CR threshold of 75.25, and they also obtain slightly higher MCR and comparable CCR. Therefore, from Fig.~\ref{fig:all_results_4a1s} and Table~\ref{tab:performance_metrics_4a1s}, one can conclude that a quantum-enhanced MAPPO solution is better suited for the 4A1S scenario than a classical one that employs the same number of parameters, especially with regards to the convergence speed, as understood in this paper.

However, the hierarchy of suitability between solutions is not the same for the 5A2S scenario. In this case, the identity-scaled architectures are always faster in terms of CS than the classical ones, but the inverse tangent-scaled ones can, at times, perform worse than the classical methods. For example, the QMARL solution of three layers and no input scaling needs slightly more time steps than the MARL solution to reach the MCR threshold of 106.1 established for the CS metric. 

The scalability of the VQC-based solution in both scenarios can be seen in Fig.~\ref{fig:all_results_4a1s} and in Fig.~\ref{fig:all_results_5a2s}. Both for the identity-postprocessing solutions and the inverse tangent-postprocessing solutions, as we increased the number of reuploading layers, the CS of each architecture always decreased, while the MCR, and the CR increased or remained at a comparable value. For the 5A2S scenario, in Table~\ref{tab:performance_metrics_5a2s}, the CCR slightly scales up with the size of the solutions, but at no statistically significant rate.

No clear correlations could be drawn when one compares the quantum metrics of the VQCs with the performance of the solutions they are embedded in. Despite having lower entanglement and expressibility values than the architectures where no input scaling is applied, the inverse-tangent scaled solutions performed better in terms of CS on the 4A1S scenario. As the number of circuit layers increases for the HQC solutions, the entanglement is reduced or stays constant, while the expressibility follows no clear path. Therefore, it is not clear if the entanglement capability or the expressibility measures could provide hints towards the scaling capabilities of QMARL solutions.

\section{Conclusion}
In this paper we introduced an aerial communication use case, in which aircrafts need to choose which communication links to create such that all aircrafts which fulfill the physical constraints are connected to base stations on the ground. Furthermore, we proposed a novel quantum-enhanced multi-agent proximal policy optimization algorithm, in which the core of the centralized critic is implemented as a variational quantum circuit, which makes use of data reuploading and of a second-order data embedding scheme. Results show that the quantum-enhanced solution outperforms the classical one in terms of maximal reward achieved at evaluation and of the convergence speed, in number of training time steps. Nevertheless, the fact that we could not draw the same empirical correlations between the QMARL solutions for the two scenarios of different complexities is an argument towards the idea that quantum-enhanced solutions need to be constructed and adapted to the specific use case they are to be applied on. Furthermore, we attempted to apply quantum architectural metrics, such as expressibility and entanglement, in order to correlate performance to the architectural properties of the quantum circuit. However, there were no clear correlations present.

Future work on this topic could include scaling the solution to a more complex and realistic use case, as well as applying other quantum architectures and compare suitability to the task. Furthermore, all results in this paper are obtained in a classical simulation of a quantum system. Therefore, a possible development branch would be to deploy this solution on quantum hardware and observe the effect of the characteristic noise and decoherence on the performance of the solution. Finally, it remains an open question and task to develop quantum architectural metrics that would offer an insight into the suitability of a quantum-enhanced solution for a given task.

\bibliographystyle{unsrt}
\bibliography{references}

\end{document}